\documentclass[12pt]{article}
\title{ Resolution of the pion puzzle:
the QCD string in Nambu-Goldstone mesons}
\author{Yu.A.Simonov\\
 State Research
Center\\Institute of Theoretical and Experimental Physics, \\
Moscow, 117218 Russia}
 \date{}
\newcommand{\beq}{\begin{eqnarray}}
 \newcommand{\eeq}{\end{eqnarray}}
\newcommand{\be}{\begin{equation}}
 \newcommand{\ee}{\end{equation}}

\def\fun#1#2{\lower3.6pt\vbox{\baselineskip0pt\lineskip.9pt
\ialign{$\mathsurround=0pt#1\hfil ##\hfil$\crcr#2\crcr\sim\crcr}}}

\newcommand{{\SD}}{\rm SD}

\newcommand{\vex}{\mbox{\boldmath${\rm x}$}}

\newcommand{\ver}{\mbox{\boldmath${\rm r}$}}

\newcommand{\lan}{\langle}
\newcommand{\ran}{\rangle}
\begin{document}
\maketitle

\begin{abstract}
Pions and kaons have double nature: the chiral dynamics  of
Nambu-Goldstone bosons together with  the usual string dynamics
common to all mesons. To uncover the interplay of both  dynamics
the effective chiral Lagrangian is derived from the QCD Lagrangian
using the Field Correlator Method and the pion self-energy (mass)
operator is written explicitly. The latter contains an infinite
number of poles, but normalized to zero at zero momentum because
of spontaneous chiral symmetry breaking. As a result one obtains
the Gell-Mann-Oakes-Renner relation for the  ground state pion and
(slightly shifted by chiral dynamics) the usual spectrum of
radially excited pions starting with $\pi(1300)$.

\end{abstract}

\section{Introduction}

The basic puzzle of the pion is that it is massless in the chiral
limit as a Nambu-Goldstone (NG) boson, while it contains
necessarily (because of confinement) a piece of the QCD string,
weighting 1 GeV per 1 fm, i.e. around 0.6 GeV for its size. A
naive quark model tells that pion differs  from the $\rho$ meson
only by spin-spin interaction, and a natural mass estimate,
neglecting the  NG mechanism, would give $m_\pi\approx 0.4$ GeV. A
similar consideration with some larger (by 200-300 MeV) resulting
mass holds also for kaon. A question arises: what is the explicit
cancellation mechanism which provides the vanishing pion mass (in
the chiral limit).

To get a hint to the solution of the puzzle one may consider the
NJL \cite{1} or the instanton model  \cite{2}, where the pion mass
indeed vanishes in the chiral limit, while the constituent quark
mass occurs due to the specific four-quark interaction.

However in both models confinement is  absent and in a sense the
puzzle stated in  the title also disappears.  One can reformulate
the problem with the Nambu-Goldstone mechanism saying, that in NJL
and instanton model the natural scale of meson mass is the twice
the  constituent mass, i.e. around 0.6 GeV, while the NG meson in
the pion channel has zero mass. Indeed in both models one can use
relations for the constituent mass to prove that the denominator
of the pion Green's function vanishes at zero momentum.

The situation in QCD is more  complicated, however. The four-quark
interaction, basic for  the spontaneous Chiral Symmetry Breaking
(CSB), occurs in QCD due to confinement (i.e. due to the  presence
of specific gluon-field correlators) and  is associated with the
QCD string \cite{3},\cite{4}. Therefore all  quarks and antiquarks
are connected by the strings and the notion of a free quark with
the constituent mass is not possible in QCD. Nevertheless
 one can do the bosonization of  the four-quark term and as a
 result one  obtains effective scalar and pseudoscalar fields from
 bilinear $q\bar q$ combinations \cite{3}.

The specific feature of the confining QCD phase is that the scalar
field condensate is in the QCD  string (scalar confining
interaction in the  effective operator), and defines quark mass
operator $\hat M$.

The outcome of this approach is the quark-meson effective
Lagrangian $L_{QML}$ and the effective  chiral Lagrangian
$L_{ECL}$, containing  only effective meson fields. In \cite{3}
these Lagrangians have been derived in the simplifying situation
when light quarks are moving in the field of the heavy antiquark.
Below we derive $L_{QML} $ and $L_{ECL}$ in  a most general
situation without heavy sources. Moreover we expand $L_{ECL}$ in
powers of pionic fields and demonstrate
 the vanishing of the pion mass in the chiral limit due to
 cancellation of two terms in the Green's function. For nonzero
 quark masses we recover the Gell-Mann-Oakes-Renner relation (GOR)
 for the pion mass.

 To study the interrelation between the quark and chiral degrees
 of freedom, the total Greens function is written in the PS
 channel which contains both nonchiral meson poles (as obtained
 e.g. from the QCD string Hamitonian \cite{5}, or quark potential models)
 and the pion chiral propagator poles.
  It is shown that due to the interplay between both types of
  poles  the lowest $q\bar q$  nonchiral pole is replaced by the
   chiral pion pole, while all others are slightly shifted downwards ("the chiral shift")
   starting with the
  first radially excited pion state. In this nontrivial and remarkable
  way one recovers the experimental spectrum in the whole pion
  channel.

  The paper is organized as follows. In the next section we derive
  $L_{QML}$ and $L_{ECL}$ following the  procedure outlined in
  \cite{3}.  In section 3 the pion quadratic terms in $L_{ECL}$
  are discussed and GOR relations are derived. In section 4 the
  $q\bar q$ Green's function is written in the PS channel and its
  pole structure is investigated in detail and compared to the
  experimental spectrum of the pion and its radial excitations.
In section 5 the situation in the pion channel is  compared to the
vector  channel, while  section 6 contains discussion and
possible development of this approach.

\section{ Effective quark Lagrangian}

 As was discussed in the previous section, one can obtain effective
 quark Lagrangian by averaging over background gluonic fields. We
 shall repeat this procedure  following \cite{3,4} now paying special
 attention to the case of light quarks.
  The QCD partition function for
 quarks and gluons can be written as
 \be
 Z=\int DAD\psi D\psi^+ {\rm exp} [L_0+L_1+L_{{\rm int}}]\label{1}
 \ee
  where we are using Euclidean metric and define
  \be
  L_0=-\frac14\int d^4x(F^a_{\mu\nu})^2,\label{2}
  \ee
  \be
  L_1=-i\int~^f\psi^+(x)(\hat
  \partial+m_f)~^f\psi(x)d^4x,\label{3}
  \ee
  \be
  L_{\rm int}=\int~^f\psi^+(x) g\hat A(x)~^f\psi(x)d^4x.\label{4}
  \ee

  Here and in what follows $~^f\psi_{a\alpha}$ denotes quark operator
  with flavour $f$, color $a$ and bispinor index $\alpha$.

  To express $A_\mu(x) $ through $F_{\mu\nu}$ one can use the
  generalized contour gauge  \cite{6} with the contour $C(x)$ from
  the point $x$ to $x_0$, which can also lie at infinity,
  \be
  A_\mu(x)=\int_C F_{\lambda\beta} (z)
  \frac{\partial z_\beta (s,x)}{\partial x_\mu}
  \frac{\partial z_\lambda}{\partial s}
  ds. \label{5}
  \ee
   Now one can integrate out gluonic field $A_\mu(x)$.
    One obtains
  \be
  Z=\int D\psi D\psi^+{\rm  exp} \{L_1+L_{\rm eff}\}\label{6}
 \ee
  where the
  effective quark Lagrangian $L_{\rm eff}$ is defined as
  \be
   {\rm
  exp} L_{\rm eff}=\langle {\rm exp} \int~^f\psi^+\hat A~^f\psi
  d^4x\rangle_A.\label{7}
   \ee
    Using cluster expansion $L_{\rm eff}$ can be written
    as an infinite sum containing averages $\langle (\hat
   A)^k\rangle_A$.  At this point one can exploit the Gaussian
  approximation, neglecting all correlators $\langle (\hat
   A)^k\rangle$ for degrees  higher than $k=2$.  Numerical accuracy
   of this approximation was discussed in \cite{7} and tested in \cite{8}.
   One expects that for static quarks corrections to Gaussian
   approximation amount to  less than 1\%.

   The resulting effective Lagrangian is quartic in $\psi$,
   \be
   L^{(4)}_{\rm eff}=\frac{1}{2N_c} \int d^4x
   d^4y^f\psi^+_{a\alpha}(x)~^f\psi_{b\beta}(x)~^g\psi^+_{b\gamma}(y)
   ~^g\psi_{a\delta}(y)J_{\alpha\beta;\gamma\delta}(x,y)+O(\psi^6),\label{8}
   \ee
   \be
   J_{\alpha\beta,\gamma\delta}(x,y)=(\gamma_{\mu})_{\alpha\beta}
   (\gamma_\nu)_{\gamma\delta} J_{\mu\nu}(x,y)\label{9}
   \ee
   and $J_{\mu\nu}$ is expressed as
   \be
   J_{\mu\nu} (x,y)=g^2\int^x_C\frac{\partial
   u_\omega}{\partial x_\mu} du_\varepsilon \int^y_C \frac{\partial
   v_{\omega'}}{\partial  y_\nu}
   dv_{\varepsilon'}\frac{tr}{N_c}\langle
   F_{\varepsilon\omega}(u)
   F_{\varepsilon'\omega'}(v)\rangle.\label{10}
      \ee
$L_{\rm eff}$ (8) is written in the contour gauge \cite{6}.

   It can be identically rewritten in the
   gauge--invariant form if one substitutes parallel
   transporters $\Phi(x,x_0),\Phi(y, x_0)$ (identically
   equal to unity in this gauge) into (8) and (10),
   multiplying each $\psi(x)$ and $\psi(y)$ respectively
   and in (10) replacing $F(u)$ by
   $\Phi(x_0,u)F(u)\Phi(u,x_0)$ and similarly for $F(v)$.

   After that $L_{\rm eff}$ becomes gauge--invariant, but in
   general contour--dependent, if one keeps only the
   quartic term (8), and neglects all higher terms.
 The final results of this paper do not depend on the choice of
 the contour (more discussion of the contour fixation is given in
 Appendix 1 of this paper).

   Till this point we have made only one approximation --neglected
   all field correlators except the Gaussian one. Now one must do
   another approximation --assume large $N_c$ expansion and keep the
   lowest term. As was shown in \cite{4}
    this enables one to replace in
   (8) the colorless product $$~^f\psi_b(x)~^g\psi_b^+(y)=
   tr (~^f\psi(x)\Phi(x, x_0)\Phi(x_0,y)~^g\psi^+(y))
   $$
   by  the quark Green's function
   \be
   ~^f\psi_{b\beta}(x)~^g\psi^+_{b\gamma}(y)\to
   \delta_{fg}N_cS_{\beta\gamma}(x,y)\label{11}
   \ee
   and $L^{(4)}_{\rm eff}$ assumes the form
   \be
   L^{(4)}_{\rm eff}=-i\int d^4xd^4y~^f\psi^+_{a\alpha}(x)
   ~^fM_{\alpha\delta}(x,y) ~^f\psi_{a\delta}(y)\label{12}
   \ee
   where the quark mass operator is
   \be
   ~^fM_{\alpha\delta}(x,y)=-iJ_{\mu\nu}(x,y) (\gamma_{\mu}~^fS(x,y)
   \gamma_\nu)_{\alpha\delta}.\label{13}
   \ee
   From (6), (\ref{12}) it is evident that $~^fS$ satisfies the equation
   \be
   (-i\hat \partial_x-im_f)~^fS(x,y)-i\int~^fM(x,z) d^4
   z~^fS(z,y)=\delta^{(4)}(x-y).\label{14}
   \ee
   Equations (13), (14) have been first derived in \cite{4}. From (6) and
   (12) one can realize that at large $N_c$ the $q\bar q$ and $3q$
   dynamics is expressed through the quark mass operator (13), which
   should contain both confinement and CSB.

At this point it is necessary to stress the important difference
of the equations (\ref{13}), (\ref{14}) from the standard
Dyson-Schwinger equations. In contrast to the latter the mass
operator $~^fM$  and the quark Green's function $S$ are not
one-body operators, but contain the string in the kernel
$J_{\mu\nu}(x,y).$ The string extends from the midpoint of $(x,y)
$ to the point $x_0$ (the distance $(x-y)$ is of the order
 of $T_g$ - the gluon correlation length of the correlator $\lan
 F(u) F(v)\ran$ in (\ref{10}), which defines  the width of the QCD
 string \cite{9}). It is convenient to specialize the contour
 $C(x,x_0)$ of  the contour gauge (\ref{5}) in such a way that it
 passes from the point $x$ to the trajectory of the center of mass
   of a gauge-invariant system, e.g. of the $\bar q q$  meson. Then the string in the mass
   operator of the quark $M_q$ extends from $x$ to the c.m.
  position of $(\bar q q)$, while the string in $M_{\bar q}$
  extends from the position of antiquark to the c.m. position. In
  \cite{3,4} the antiquark was taken to be infinitely heavy, so
  that one could consider only one string in $M_q$.

  Nonlinear equations (\ref{13}), (\ref{14}) have been studied in
  \cite{4}  and it was shown that there is a solution with the
  scalar confining interaction for $~^fM(x,y)$, which implies
CSB, and in  \cite{4} first estimates
  were given  for the chiral condensate $\lan \bar q q\ran$, $\lan
  \bar q q\ran \sim  -c\sigma^{3/2}, ~~c= 0.5 \div 0.7$ depending
  on the nonlocality of the mass operator $~^fM$, however
  convergence of the series  defining the constant $c$ was rather
  slow.

  To discuss the NG modes and the CSB effects in more detail, we
  shall use the bosonization method of \cite{3}   in our general setting. The only difference from \cite{3}
  is that we are
  using the most general contour gauge (\ref{5}) with the contour prescription for
   the final gauge-invariant
  quantity (like Green's function), which ensures the minimal area
  worldsheet for the string and  hence for the total action of the
  system.

  Now instead of the large $N_c$ substitution (\ref{11}) one can
  do the  bosonization of the 4q Lagrangian (\ref{8}) in  the same
   way as it was done in \cite{3} for the heavy-light meson case.
   Leaving details to the Appendix 2, one can  write the resulting
   quark-meson Lagrangian for $n_f =2$ keeping only
   scalar-isoscalar and pionic effective fields as follows
$$
   L_{QML} =\int d^4x\int d^4 y \{ ~^f\psi^+_{a\alpha} (x)[ (i\hat
   \partial + i m_f)_{\alpha\beta} \delta_{fg} \delta^{(4)}(x-y) +
   i M_{\alpha\beta}^{(fg)} (x,y) ] ^g\psi_{a\beta} (y)-
$$
\be
-[\chi^{(0,S)}(x,y) \chi^{(0,S)}(y,x) + \chi_a^{(1,PS)}(x,y)
\chi_a^{1,PS)}(y,x) ]J(x,y)\}\label{15} \ee where $M$ is
\be
M_{\alpha\beta}^{(fg)} (x,y) = (\chi^{(0,S)} (x,y) \delta_{a\beta}
t^{(0)}_{fg} + \chi^{(1,PS)}_a (x,y) (i\gamma_5)_{\alpha\beta}
t^{(a)}_{fg})J(x,y)\label{16}\ee

It is convenient to use another parametrization of the pionic
field  replacing $\chi^{(0,S)}$ and  $\chi_a^{(1,PS)} $ by $M_S$
and $\phi_a$ \be M^{(fg)}_{\alpha\beta} (x,y) = M_S(x,y) \exp
(i\gamma_5 t^a\phi_a(x,y))_{\alpha\beta}^{(fg)}\equiv M_S(x,y)
\hat U(x,y). \label{17}\ee Then (\ref{15}) can be rewritten as $$
 L_{QML} =\int d^4x\int d^4 y \{ ~^f\psi^+_{a\alpha} (x)[ (i\hat
   \partial + i m_f)_{\alpha\beta} \delta^{(4)}(x-y) \delta_{fg}
   +$$
   \be
   i M_S \hat U_{\alpha\beta}^{(fg)}(x,y)]~^g\psi_{a\beta}(y)- 2n_f[J(x,y)]^{-1}M_S^2(x,y) \}.
\label{18}\ee

The partition function with $L_{QML}$ assumes the form
\be
Z=\int D\psi D\psi^+ DM_SD\phi_a \exp L_{QML}.\label{19}\ee One
can integrate over quark fields in (\ref{19}) yielding the
effective chiral Lagrangian $L_{ECL}$
\be
Z=\int D M_SD\phi_a \exp L_{ECL}\label{20}\ee where it is assumed
that $L_{ECL}$ is considered in the nonsinglet channels and
therefore the chiral anomaly contribution can be neglected.
$L_{ECL}$ has the form
\be
L_{ECL} =-2n_f\int d^4x \int d^4y (J(x,y))^{-1}M_S^2(x,y) +N_c
tr\log [(i\hat
\partial +im_f)\hat 1+ iM_S\hat U].\label{21}\ee Here $tr$ refers
to the flavour, Lorentz indices and coordinates, and $\hat 1_{fg}=
\delta_{fg}$, and the expression in the square brackets in
(\ref{21}) is the $n_f\times n_f$ matrix. To find the minimum of
$L_{ECL}$ as a functional of $M_S, \phi_a$ one obtains the
stationary point equations yielding
\be
\frac{\delta L_{ECL}}{\delta M_S(x,y)} = -4n_f(J(x,y))^{-1} M_S
(x,y) - N_c tr (S_\phi ie^{i\gamma_5\hat \phi})=0,\label{22}\ee
\be
\frac{\delta L_{ECL}}{\delta\phi_a(x,y)} =  N_c tr (S_\phi M_S
e^{i\gamma_5\hat \phi}\gamma_5 t_a)=0,\label{23}\ee where $
S_\phi=-[(i\hat \partial+ im_f){\hat 1}+iM_S\hat U]^{-1}$. The
solutions of (\ref{22}), (\ref{23}) are $\phi_a^{(0)}=0,
M_S=M_S^{(0)},$ where $M_S^{(0)}$ satisfies relation
\be
iM^{(0)}_S(x,y) =\frac{N_c}{4} tr (SJ(x,y))= (\gamma_\mu
S\gamma_\nu)_{sc} J_{\mu\nu} (x,y)\label{24}\ee and the subscript
$sc$ implies taking the Lorentz scalar part of the operator and
$S\equiv S_\phi(\phi_a=0)$. Note that eq. (\ref{24}) coincides
with (\ref{13}) if one neglects in $M_S$ , as we have done before,
all terms except scalars.

This remarkable feature demonstrates a completely new meaning of
the scalar condensate in QCD: while in nonconfining models like
NJL or instanton model the scalar condensate like  Higgs
condensate  is constant everywhere, in QCD  the scalar condensate
is concentrated in the string  - i.e. in $M_S(x,y)$ - and is
actually the dominant part of the string itself.

This is of course the consequence of the fact, that in QCD the
chiral symmetry breaking is due to confinement, and the effective
meson fields are nonzero just where the confining kernel
$M_S(x,y)$ is present.

\section{The pion mass}

In this section it will be proved that the pion is massless as it
should be for the Nambu-Goldstone meson in the chiral limit and
that for nonzero quark mass one has the Gell-Mann-Oakes-Renner
relation \cite{10}. To this end we consider the pionic part of the
$L_{ECL}$ (\ref{21}) and expand it in powers of the pionic  field
$\phi_a$ up to the second order.

One has $$-W(\phi)\equiv N_c tr \log [(i\hat \partial+im_f)\hat
1+iM_S\hat U]=$$ $$N_ctr \log [(i\hat \partial + im_f+iM_S)\hat 1
+\Delta]=$$ $$N_ctr \log ((i\hat \partial + im_f+iM_S)\hat  1)+
N_ctr [(i \partial + im_f+iM_S)^{- 1}\Delta]-$$
\be
-\frac{1}{2}N_ctr [(i\hat \partial + im_f+iM_S)^{-1}\Delta (i\hat
\partial + im_f+iM_S)^{-1}\Delta].\label{25}\ee
Here $\Delta=(-\hat \phi\gamma_5-\frac{i}{2}\hat \phi^2)M_S,~~
\hat \phi\equiv \phi_a t_a.$ One can rewrite the quadratic in
$\phi_a$ terms as
\be
W^{(2)}(\phi) =\frac12 \int \frac{d^4k}{(2\pi)^4}
\frac{d^4k'}{(2\pi)^4} \phi_a (k) \hat N(k,k')
\phi_a(k').\label{26} \ee

To simplify derivation we shall neglect the isospin violation and
nonlocality, replace $m_f\to \frac{m_u+m_d}{2} \equiv m$, and
write $M_S(x,y)\to M_S(x)\delta^{(4)}(x-y),~~ \phi_a(x,y)\to
\phi_a(x)$. In \cite{4} it was shown that $M_S(x,y)$ indeed has
this property at least at large distances,  where  also
$M_S^{(0)}(x) =\sigma|\vex|$.

From (\ref{26}), (\ref{25}) one obtains $$ N(k,k')=\frac{N_c}{2}
\int dx e^{i(k+k')x} tr (\Lambda M_S)_{xx} + \frac{N_c}{2} \int
d^{(4)}(x-y)d^4(\frac{x+y}{2})\times$$
\be
e^{\frac{i}{2}(k+k') (x+y) +\frac{i}{2} (k-k') (y-x)} tr [
\Lambda(x,y) M_S(y) \bar \Lambda (y,x) M_S(x)]\label{27}\ee where
we have defined
\be
\Lambda= (\hat \partial + m + M_S)^{-1},~~ \bar \Lambda = (\hat
\partial - m-M_S)^{-1}.\label{28}\ee
One can  use translation invariance of the traces in (\ref{27}) to
rewrite (\ref{26}) as
\be
W^{(2)}(\phi) =\frac{N_c}{2} \int \phi_a (k) \phi_a(-k) \bar N(k)
\frac{d^{(4)}k}{(2\pi)^4}\label{29}\ee with
\be
\bar N(k)= \frac12 tr\{(\Lambda M_S)_0+ \int
d^{(4)}ze^{ikz}\Lambda(0,z) M_S(z) \bar \Lambda(z,0)
M_S(0)\}.\label{30}\ee The pion mass is proportional to $\bar
N(0)$, which can be written after some algebraic manipulations as
(see Appendix 3 for details of derivation).
\be
\bar N(0) =\frac12 tr (\Lambda M_S\bar \Lambda (\hat \partial - m)
)= \frac{m}{4} tr (\Lambda-\bar \Lambda) = \frac{1}{2}m tr \Lambda
+ O(m^2) .\label{31}\ee Note here that in the chiral limit $m\to
0$, one has $m^2_\pi\sim N(0,0)=0$. In all calculations resulting
in (\ref{30}), (\ref{31}) the use is done of the reflection
symmetry, which allows to replace $(-\hat \partial)$ by $(\hat
\partial)$. A similar calculation was done earlier in \cite{11}
where instantons have been used to create the four-quark operators
in the background of gluon fields, but the phenomenon of the  CSB
due to confinement was absent.

Since the quark condensate as defined in the Minkowskian
space-time $\lan \psi\bar \psi\ran_M$ is connected to $tr \Lambda$
as
\be
\lan \psi \bar \psi\ran_M=i \lan \psi\psi^+\ran_E=i N_c tr S (x,y)
= -N_c tr \Lambda\label{32}\ee one has for the pion mass with the
usual normalization   $\phi_a=\frac{2\pi_a}{f_\pi}, f_\pi = 93
MeV,$\be 2m^2_\pi f^2_\pi = (m_u +m_d) | \lan \psi \bar \psi
\ran_M|= (m_u+ m_d)|\lan\bar u u+\bar d d\ran|\label{33}\ee which
in the standard Gell-Mann-Oakes-Renner relation \cite{10}.

\section{The pion Green's function}

To obtain the pion Green's function one can use the partition
function (\ref{19}) to calculate the correlator of the PS currents
$J_a^{(5)}(x) = \psi^+ (x) \gamma_5 t_a\psi(x)$,
\be
G_{ab} (x,y) =\frac{1}{Z} \int D\psi D\psi^+ DM_S D\phi \exp
(L_{QML}^{(2)}) J^{(5)}_a(x) J^{(5)}_b(y).\label{34}\ee
Integrating over $D\psi D\psi^+$ one gets the standard expression
$$ G_{ab} (x,y) =\frac{1}{Z} \int  DM_S D\phi e^{L_{ECL}^{(2)}}\{
 Tr [ S_\phi (x,y) \gamma_5 t_b S_\phi (y,x) \gamma_5 t_a]-$$
 \be
 - Tr (S_\phi(x,x) \gamma_5 t_a) Tr (S_\phi(y,y) \gamma_5
 t_b)\}\label{35}\ee
 where $S(x,y)$ is defined in (\ref{23}) and depends on $M_S$ and
 $\phi_a$. Our following discussion  is similar  in some respect to  the
 line of reasoning in \cite{11},
 however instantons are omitted in (\ref{35}) since
  the CSB occurs due to confinement, as it is
 demonstrated above explicitly by Eq. (\ref{24}).

Eq. (\ref{35}) contains two terms with one and two trace operators
in the curly brackets, which can be called the connected and  and
the disconnected terms respectively. The integration over $DM_S$
can be done using the stationary condition (\ref{24}), while for
the integration in $D\phi$ one expands $L_{ECL}^{(2)}$ in $\phi_a$
around the stationary point $\phi_a^{(0)}=0$ keeping the second
order terms. In what follows $M_S$ will be replaced by  the
stationary point value $M_S^{(0)}$ without changing notation.
Using now the expansion for $S_\phi$
\be
S_\phi(x,y) = S(x,y) + S(x,z) M_S \gamma_5 \hat \phi (z) S(z,y)
+O(\phi^2)\label{36}\ee and noticing that $\phi_a=2\pi_a/f_\pi$
and $f_\pi= O(N_c^{1/2})$, one can deduce that the emission or
absorption of the pion contains  a factor $N_c^{-1/2}$, or in
other words, the coupling constant $ g_{\pi q\bar q} =
O(N_c^{-1/2})$. Therefore in the large $N_c$ limit one should take
into account the lowest number of pion exchanges and we shall
neglect pion fields in the connected term in (\ref{35}), replacing
$S_\phi$ by $S\equiv S_\phi(\phi=0)$,
 and keeping only the first order term (\ref{36}) in the
 disconnected term of (\ref{35}). Integrating now over $D\phi$,
 one obtains from both terms the contribution, proportional to
 $N_c$
 \be G_{ab} (x,y)= N_cG_{ab}^{(0)} (x,y) -
 \frac{N^2_c}{f^2_\pi}G^{(M)}_{ac} (x,z) G^{(0)}_{\pi} (z,u)
 G_{cb}^{(M)}(u,y) \label{37}\ee
 where we have defined
 \be
  G_{ab}^{(0)} (x,y) \equiv \frac12 G^{(0)}\delta_{ab} \equiv tr [
  S(x,y) \gamma_5  t_b S(y,x) \gamma_5 t_a]\label{38}\ee
\be
 G_{ab}^{(M)} (x,y) \equiv \frac12  G^{(M)}\delta_{ab}
\equiv tr [ \gamma_5 t_a   S(x,y) M_S\gamma_5  t_b
S(y,x)]\label{39}\ee
\be
G^{(0)}_\pi (x,y)= \int \frac{d^4k}{(2\pi)^4}
\frac{e^{ik(x-y)}f^2_\pi}{ N_c \bar N(k)},\label{40}\ee and the
sign $tr$ implies Lorentz, flavour and coordinate summation
(integration).

Therefore the total Green's function looks like
\be
G(x,y) =N_c \{ G^{(0)} (x,y) -\frac{N_c}{2f^2_\pi} G^{(M)} (x,z)
G^{(0)}_\pi (z,u) G^{(M)}(u,y)\}.\label{41}\ee Using (\ref{30})
and going over to the momentum space one has \be G(k) = N_c\{
G^{(0)} (k) - \frac{ G^{(M)}(k) \cdot G^{(M)}(k)}
{ G^{(MM)}(k) +tr (\Lambda M_S)}\}.\label{42}\ee Here
$G^{(MM)}(k) \equiv tr (SM_S\gamma_5SM_S\gamma_5)_{k }$, where the
subscript $k$ implies the Fourier transform which according to
(\ref{31}) is
\be
\bar N (k) =\frac12(G^{(MM)} (k)+tr (\Lambda M_S)) = (m^2_\pi +
k^2) \frac{f^2_\pi}{4N_c} + 0(k^4).\label{43}\ee

Therefore (\ref{42}) contains the pion pole in the second term on
the r.h.s. of (\ref{42}).

The question arises what happens with the  poles of $G^{(0)}( k)$.
Actually here appears the  lowest pole in the PS channel, which
one can  call the   Quark Model (QM) pion, usually situated around
400 MeV. This pole is present in all three Green's functions
$G^{(0)} (k), G^{(M)} (k)$ and $G^{(MM)}(k)$, and if one contracts
all $M_S$ factors in the numerator and denominator of (\ref{42}),
one can easily see that the QM pole is cancelled between the
first  and the second term in the curly brackets.

To make this cancellation more transparent one  can again use the
large $N_c$ argument  and to represent $G^{(0)}(k), G^{(M)} (k)$
and $G^{(MM)}(k)$ as series of poles, e.g.
\be
G^{(0)}(k)=-\sum^\infty_{n=0} \frac{c^2_n}{k^2+m^2_n},~~
G^{(M)} (k)=-\sum^\infty_{n=0} \frac{c_n c_n^{(M)}}{k^2+m^2_n},~~
G^{(MM)}(k)=-\sum^\infty_{n=0}
\frac{(c_n^{(M)})^2}{k^2+m^2_n}\label{44}\ee where $c_n^{(M)}$
differs from $c_n$ by the presence of the operator $M$ inside the
matrix element, at the initial or final point of the Green's
function, e.g. for $G^{(MM)} (k)$
\be
G^{(MM)} (k)= \int tr (\gamma_5 M_S(x) S(x,y) \gamma_5 M_S(y)
S(y,x)) e^{ik(x-y)}  d(x-y).\label{45}\ee The presence of the
scalar quasilocal in time operator $M_S(x)$ cannot change the
spectrum of  bound states in  $G^{(M)} (k)$ and $G^{(MM)}(k)$
as  compared to $G^{(0)} (k)$ where operators $M_S(x)$
are absent, and  therefore all three functions should have the
same set of poles.

Now the vanishing of $\bar N(0)$ for $m=0$ implies that in the
chiral limit one can write the denominator in (\ref{42}) as
\be
G^{(MM)}(k) + tr \Lambda M_S = G^{(MM)} (k) -G^{(MM)} (0) = k^2
\sum^\infty_{n=0} \frac{(c_n^{(M)})^2}{(m^2_n+ k^2)m^2_n }.
\label{46}\ee

Hence the total Green's function can be written as (in the chiral
limit, $m\to 0$). \be G(k) =\frac{\Psi(k)}{k^2 \Phi(k)},
\label{47}\ee where
\be
\Psi(k) = \sum^\infty_{n=0} \frac{c_n}{k^2+m^2_n}\sum^\infty_{k=0}
\frac{[c_n(c^{(M)}_k)^2\frac{k^2}{m^2_k} + c_n^{(M)} c_k
c_k^{(M)}]}{k^2+m^2_k},\label{48} \ee
\be
\Phi(k)=-\sum^\infty_{n=0} \frac{(c_n^M)^2}{m^2_n
(k^2+m^2_n)}.\label{49}\ee

From the structure of (\ref{47}) one can easily see, that double
poles at $k^2=-m^2_n$ are cancelled in $\Psi(k)$, so that only
simple poles are retained, which are compensated by the same poles
of $\Phi(k)$, so that the ratio $\Psi(k)/\Phi(k)$ does not have
poles at $k^2=-m^2_n, n=0,1,2,...$

Clearly $G(k)$ has a pole at  $k^2=0$, which is the expected pion
pole of the chiral limit, which is shifted to the   position
$k^2=-m^2_\pi$, with $m^2_\pi$ defined in (\ref{33}). Now the
question arises  where  appears the next pole in $G(k)$.

To this end one can take into account the properties of the hadron
spectrum in the linearly confining vacuum, found  in the vector
channel in \cite{12}, and in general in \cite{5,13,14}, which are
obtained without chiral  effects, i.e. exactly pertinent  to $m_n,
c_n^{(M)}$. It  was found that the spectrum of radially excited
mesons in the large $N_c$ limit satisfies approximate relation
\be
m^2_n\simeq 4\pi \sigma n + \Delta m, ~~ \Delta m
(J^{PC}=0^{-+})\simeq 0.4 {\rm GeV}\label{50}\ee while $c_n^{(M)}$
is roughly independent of $n$. The latter is due to the fact that
$c_n^{(M)}\sim \psi_n (\ver =0)$ and for the linear interaction
$\psi_n(0)$ does not depend on  $n$.

Using  these properties one can find  zeros of $\Phi(k)$ in the
$k^2$ plane $k^2=(k^{(0)})^2$, which correspond to the poles in
$G(k)$. One has from (\ref{49})
\be
(k^{(0)})^2=- m^2_1 (1 -\delta_1), ~~ -m^2_2(1-\delta_2),... -
m^2_n(1-\delta_n)...\label{51}\ee where   the chiral shifts of the
levels are:  $\delta_1\simeq
\frac{c^2_1m^2_0(m^2_1-m^2_0)}{m_1^2(c_0^2 m_1^2+c_1^2m_0^2)},~~
\delta_2 \simeq O(m^2_0),$ etc. Hence the first excited pion pole
is close to the position of the first excited pole obtained in the
quark model approach without taking into account chiral effects.
The same is true for higher excited states. It is interesting that
the ground QM state with mass $m_0\approx 0.4$ GeV disappears from
the total spectrum of the full Green's function $G(k)$, and is
replaced by the correct chiral-generated pole at $m_\pi$  given in
(\ref{33}). This is exactly what one should expect on physical
grounds and it is rewarding that our simple equation (\ref{42})
reproduces this physically reasonable result.

The calculation of the radially excited states in the framework of
the QCD string approach was recently done in \cite{14}. For the
first radially excited state the $1/N_c$ corrections are not large
and the spin-averaged mass was found to be  $\bar M_{av} = 1.45$
GeV with $ \sigma=0.18$ GeV$^2$  (the only input). The pion mass
is shifted downwards by roughly 0.105 GeV due to hyperfine
interaction and the resulting theoretical pion mass $m_1$ is
around 1.35 GeV, which  after subtraction of the chiral shift
$\delta_1$ exactly coincides with  the experimental mass value of
$\pi(1300)$.

\section{Discussion: photon and pion selfenergy operators}

The pion Green's function
\be
G^{(0)}_\pi(k) \sim \frac{1}{N(k)} \sim \frac{1}
{G^{(MM)}(k)-G^{(MM)}(0)}\label{52}\ee which is obtained from the
effective Lagrangian $L_{ECL}$, Eq. (\ref{21}), contains an
infinite number of poles in the denominator due to the self-energy
operator, the role of the latter being played by
 $G^{(MM)}(k)$. These poles are due to confinement only
 while chiral effects (pion emission and absorption) are switched
 off in $G^{(MM)}(k)$ (hence the superscript (0)).

 Now by construction -- and this is the consequence of the
 spontaneous CSB, in the pion Green's function only the difference
  $G^{(MM)}(k)-G^{(MM)}(0)$ enters in the denominator
  which is $O(k^2)$ in the chiral limit $(m=0)$ for $k\to 0$.

  The situation here is  similar to the case of
  the photon propagator with the self-energy operator due to the
  hadron intermediate states, as it enters in the process
  $e^+e^-  \to $ hadrons. Indeed the full transverse photon Green's
  function $D(k^2)$ can be  written  as
  \be
  D(k^2)= \frac{4\pi}{k^2-\mathcal{P}(k^2)},~~
  \mathcal{P}_{\mu\nu} (k) = (g_{\mu\nu}- \frac{k_\mu
  k_\nu}{k^2})\mathcal{P}(k^2)\label{53}\ee
  with the renormalization conditions
  $\mathcal{P}(k^2=0)=\mathcal{P'}(k^2=0)=0$.
  $ \mathcal{P}(k^2)$ is connected to the  vacuum polarization
  operator $\Pi(k^2)$ via $ \mathcal{P}(k^2) \sim e^2 k^2
  \Pi(k^2)$,  and $Jm \Pi (k^2)$ is proportional to the hadronic ratio $R(k^2)$ of the process
  $e^+e^- \to $ hadrons.

  Using the same reasoning as in \cite{12}, one can express for
  large $N_c$ the operator $\Pi(k^2)$ as a sum over hadronic poles
  - radial excitations in the vector channel with masses $m_n(v)$,
  $\Pi(k^2) =\sum\frac{a_n}{k^2+m^2_n(v)}$. As a result, the
  structure of the photon Green's function (\ref{53}) becomes
  similar to that of the pion Green's function (\ref{52}): an
  infinite sum of poles in the denominator renormalized by
  subtractions  to vanish for $k=0$. The latter condition comes
  from the spontaneous CSB (the NG theorem) in case of  the  pion and
  from gauge invariance  in case of  the photon. It seems   that one
  could use the same Eq. (\ref{42}) for both cases with similar
  conclusions for the spectrum.

  However at this point the similarity stops. Indeed, in case of
  the photon, the mixing between photon and vector meson states is
  governed by $\alpha=\frac{1}{137}$ and is small, and the  number
  of states does not change, it consists of all vector meson
  states plus photon.  In this case the vector meson states are
  shifted only slightly as can be seen from (\ref{53}) yielding
  condition for poles; $\Pi(k^2)=\frac{const}{e^2} $ and  as a
  result $k^2_n=(k^{(0)}_n)^2+d_ne^2$.

  In case of the pion, the situation is different; the pion and
  hadron resonances in the self-energy part are not coupled by a
  small coupling constant, but rather enter in the denominator
  with the large coefficient, such that it exactly cancels the
  lowest hadron state trading it for the chiral pion. Therefore
  the total spectrum consists  of  the almost massless pion and
  all its radial excitations, while the lowest nonchiral  state
  disappears from the spectrum. This is rewarding since the chiral
  pion is made of the same  $\bar q q$ degrees of freedom as the
  nonchiral states, and the number of states should not increase
  because part of states is treated differently.

This is not true for the photon, since the  photon has  a
different
  nature and origin as compared to hadron states in $\Pi(k^2)$.

\section{Conclusions and outlook}

The  purpose of this paper is twofold. First of all, it is
necessary to understand for the system of light quarks how
confinement spontaneously provides the chiral symmetry breaking.
For the heavy-light system this was first shown in \cite{4} and
developed  making use of Green's functions and Dirac formalism.
Later in \cite{3}
 the same connection between confinement and CSB was established
 using the familiar bosonization technic, but again for the
 heavy-light systems. In the present paper the bosonization and
 the resulting effective chiral Lagrangian is obtained for the
 light-light systems in Eqs. (\ref{18}) -(\ref{21}). It  is
 remarkable, that the stationary point  equations (\ref{22}),
 (\ref{23}) yield a nonlinear equation for the mass operator
 (\ref{24}), with the self-consistent  solution (discussed already
 in \cite{4}) containing the scalar string between quark and
 antiquark. This is similar to the NJL and the instanton models
 where in absense of confinement  also the stationary point
  solution brings about the scalar
 condensate - and this appears spontaneously implying CSB. The
 difference is that in case of confinement this scalar condensate
 resides  not in all space but in strings and also gives CSB.

 Another important purpose of the paper is to clarify the problem
 of the  pion mass: while for all other mesons the QCD string approach
 and also the constituent quark model give a reasonable
 description of  the spectrum, the spectrum of pions  (and also of
 kaons and $\eta,\eta')$ is strikingly different, with the very
 large mass gap between $\pi(139)$ and $\pi(1300)$. In the present
 paper we have derived the total Green's function in the PS
 channel, containing both chiral and confining properties. It was
 shown for the first time that the effective chiral Lagrangian
 derived directly from the QCD Lagrangian produces  the pion
 mass satisfying the Gell-Mann-Oakes-Renner relation. Moreover
 it was demonstrated that poles of the total Green's function
 correspond exactly  to the  physical picture: namely, the lowest
 QM pole $m^2_0$ moves into the pseudo-Nambu-Goldstone position,
 while all the higher poles shift only slightly reproducing the
 experimental picture.

 For simplicity reasons only pion was considered above. It is not
 difficult to extend the formalism to $SU(3)$ and to treat $\eta$
 and kaons, and the pion and kaon Regge-trajectories. The $\eta'$
 case includes the chiral anomaly and connection to the gluonic
 (glueball) channel and this study is planned for the future.

 Another part of the work, which is still remains not done is the
 study of the pion wave function and of the role of confinement in
 its dynamics.

The author is grateful for useful remarks to S.M.Fedorov and
N.O.Agasian.
 This work was done with the support of the grants RFBR -- 00-15-96786,
 INTAS 00-00110  and INTAS  00-00366.

\newpage

 \setcounter{equation}{0}
\renewcommand{\theequation}{A.\arabic{equation}}

\begin{center}
{\bf Appendix 1}\\

\end{center}

The gauge-invariant form of (\ref{8}), (\ref{10}) obtains by
insertion of the parallel transporters $\Phi_C(z, x_0)$ to all
$\psi$'s in (\ref{8}) and replacing $F(z)$ by $\Phi_C(x_0,z) F(z)$
$ \Phi_C(z, x_0)$. The resulting effective $4q$ Lagrangian (8)
will depend on the shape of the contour $C$. It is clear that the
whole sum of all-order correlators, $\sum^\infty_{n=4,6...}
L^{(n)}_{eff}\equiv L_{eff}$ does not depend on the shape of $C$.
So with $L_{eff}$ one can choose any contour $C(x)$ and in
particular to select a contour $C(x)$ in such  a way as to produce
the minimal area surface for the world-sheet of the string, or in
the Lagrangian formalism - to form the string (contour) of minimal
length between the quark at one end and antiquark on another end
of the string.

For this minimal string (minimal area surface) it is legitimate to
drop all terms in $L_{eff}$ except $L^{(4)}_{eff}$, since the
difference (for the minimal area), is expected to be of the order
of 1\%, according to \cite{7,8}.

    A
   similar problem occurs in the cluster expansion of
   Wilson loop, when one keeps only lowest correlators,
   leading to the (erroneous) surface dependence of the
   result. Note that only for the minimal area surface higher
   correlators are negligible.
Therefore the Gaussian approximation is valid when the minimal
area surface is fixed beforehand.
   Situation here is the same as with the  sum of QCD
   perturbation series, which depends on the
   normalization mass $\mu$ for any finite number of
   terms in the series. This unphysical dependence is
   usually treated by fixing $\mu$ at some physically
   reasonable value $\mu_0$ (of the order of the inverse
   size of the system).

Similarly, the
   physical choice of the contour corresponds to the minimization of
   the meson (baryon) mass over the class of strings, generated by
   contours $C$
   in the same way as the choice of $\mu=\mu_0$
   corresponds to the minimization of the dropped higher
   perturbative terms.

   As a practical outcome, we shall keep the set  of contours $C$ till the end
   and finally use it to minimize the string between the quarks.\\


 \setcounter{equation}{0}
\renewcommand{\theequation}{A.\arabic{equation}}

\begin{center}
{\bf Appendix 2 }\\

 \vspace{0.5cm}

{\large Derivation of the effective quark-meson Lagrangian
$L_{QML}^{(2)}$}
\end{center}

We start with the white quark bilinears, as in \cite{3}
\be
\Psi^{fg}_{\alpha\varepsilon}(x,y)\equiv ~^f\psi^+_{a\alpha}(x)
~^g\psi_{a\varepsilon}(y)= ~^f\psi^+_{a'\alpha} (x)
\Phi_{a'a}(x,Y) \Phi_{ac}(Y,y)~^g\psi_{c\varepsilon}(y)
\label{A.1}\ee and introduce the   isospin generators
$t^{(n)}_{fg}$
\be
\sum^{n^2_f-1}_{n=0}t^{(n)}_{fg}t^{(n)}_{ij}= \frac12
\delta_{fj}\delta_{gi}; t^{(0)}=\frac{1}{\sqrt{2n_f}}\hat 1, ~~
tr(t^{(n)}t^{(k)}) =\frac12 \delta_{nk}. \label{A.2} \ee

Hence the bilinears in (\ref{8}) can be written as
\be
\Psi^{fg}_{\alpha\varepsilon}(x,y) \Psi^{gf}_{\gamma\beta} (y,x) =
2 \sum^{n^2_f-1}_{n=0} \Psi^{(n)}_{\alpha\varepsilon}(x,y)
\Psi^{(n)}_{\gamma\beta} (y,x) \label{A.3} \ee where we have
defined
\be
\Psi^{(n)}_{\alpha\varepsilon}
 (x,y) \equiv ~^f\psi^+_{a\alpha}(x)t^{(n)}_{fg}~^g\psi_{a\varepsilon}(y).
\label{A.4} \ee Now one can use the Fierz transformation for the
combination $\gamma^\mu\gamma^\nu$  (see Appendix 3 of \cite{3})
\be
(\gamma^{\mu})_{\alpha\beta} (\gamma^{\nu})_{\gamma\delta}=
\frac14\sum^A\Delta_A
(\gamma^\mu\gamma_A\gamma^\nu)_{\alpha\delta}
(\gamma_A)_{\gamma\beta}\label{A.5}\ee where $\Delta_A=-1$ for
$\gamma_A=\gamma_5 \gamma_\mu$ and $\Delta_A=1$ otherwise. We
shall be interested here only in scalar and pseudoscalar
combinations on the r.h.s. of (\ref{A.5}), hence one can write
\be
(\gamma^{\mu})_{\alpha\beta} (\gamma^{\nu})_{\gamma\delta}=
\frac14\delta_{\mu\nu} \{1_{\alpha\delta} 1_{\gamma\beta} +
(i\gamma_5)_{\alpha\delta} (i\gamma_5)_{\gamma\beta} \} +...
\equiv\frac14 \delta_{\mu\nu} \{O_{\alpha\delta}^{(S)}
O_{\gamma\beta}^{(S)}+O_{\alpha\delta}^{(PS)}
O_{\gamma\beta}^{(PS)}\}...\label{A.6}\ee where ellipsis in
(\ref{A.6}) and in what follows implies contribution of all other
combinations. As the result one has for $L_{eff}^{(4)}$, Eq.
(\ref{8}) \be L_{eff}^{(4)} =-\int d^4x\int
d^4y\{\Psi^{(n,S)}(x,y)\Psi^{(n,S)}(y,x)+
\Psi^{(n,PS)}(x,y)\Psi^{(n,PS)}(y,x)\}J(x,y)+...\label{A.7}\ee
where $J(x,y)\equiv \frac{1}{N_c} J_{\mu\mu}(x,y)$ and
\be
\Psi^{(n,S),(n,PS)} (x,y)=\frac12\Psi^{(n)}_{\alpha\beta}
(1_{\alpha\beta}, (i\gamma_5)_{\alpha\beta}).\label{A.8}\ee

Now the Hubbard-Stratonovich transformation is written
symbolically through effective nonlocal bosonic fields
$\chi^{(n,S)}(x,y), \chi^{(n,PS)}(x,y)$ as follows
\be
e^{-\Psi\tilde J\Psi}= \int(\det J)^{1/2} D\chi\exp (-\chi J\chi+
i\Psi J \chi + i\chi  J \Psi) \label{A.9} \ee and  the partition
function  assumes the form
\be
Z=\int D\psi D\psi^+ D\chi \exp L_{QML} \label{A.10} \ee where the
effective quark-meson Lagrangian is $$
 L^{(2)}_{QML}
=\int d^4x\int
d^4y\{~^f\psi^+_{a\alpha}(x)[(i\hat\partial+im_f)_{\alpha\beta}\delta_{fg}\delta^{(4)}(x-y)
+iM^{(fg)}_{\alpha\beta} (x,y)]~^g\psi_{a\beta}(y)- $$\be
-\sum_{k=S,PS}\chi^{(n,k)}(x,y) J(x,y) \chi^{(n,k)}(y,x)\}
\label{A.11} \ee
 and the
effective quark-mass operator is
\be
M^{(fg)}_{\alpha\beta}(x,y) =\sum_{n=0,... n^2_f-1,k=S,PS}
\chi^{(n,k)}(x,y) O^{(k)}_{\alpha\beta}t^{(n)}_{fg}J(x,y).
\label{A.12} \ee\\


\begin{center}
{\bf Appendix 3 }\\
 \vspace{0.5cm}

 {\large Derivation of the GOR
relation(\ref{33}) }
\end{center}

  We start with the definition (\ref{30}) which can be rewritten
  identically as
  $$
  \bar N(0)= \frac12 tr [ \Lambda M_S\bar \Lambda(\hat \partial
  -m- M_S) + \Lambda  M_S\bar \Lambda M_S] =
$$
\be
=\frac12 tr [\Lambda M_S\bar \Lambda (\hat \partial
-m)].\label{A.13}\ee Inserting for the factor $M_S$ in the last
equality of (\ref{A.13}) the expression
\be
M_S=\frac12 (\Lambda^{-1} -\bar \Lambda^{-1} -2m),\label{A.14}\ee
one immediately obtains
\be
 \bar N(0)= \frac12 tr (A+B+C),\label{A.15}
 \ee
 where
 $$
 A\equiv \frac12 (\bar \Lambda- \Lambda)\hat \partial,~~
  B\equiv
 \frac{m}{2} (\Lambda-\bar \Lambda),~~
C=-m\Lambda \bar \Lambda(\hat \partial -m).$$

It is easy to see, that  $A$  vanishes since it is odd with
respect to  the  space-time reflection, and $C$ is $O(m^2)$. The
term $B$ yields the answer given in Eq.(\ref{31}).


\begin{thebibliography}{99}

\bibitem{1} Y.Nambu and  G.Jona-Lasinio, Phys. Rev.  {\bf 122},345
(1961);\\S.P.Klevansky, Rev. Mod. Phys. {\bf 64}, 650 (1992).

\bibitem{2}
D.Diakonov and  V.Petrov, Nucl. Phys. {\bf B272}, 457 (1986); for
a review and earlier refs. see Th.Shaefer and E.V.Shuryak, Rev.
Mod. Phys. {\bf 70}, 323 (1998).

\bibitem{3}
Yu.A.Simonov, Phys. Rev. {\bf D65}, 094018 (2002),
hep-ph/0201170.

\bibitem{4}
Yu.A.Simonov, Phys. Atom. Nucl {\bf 60}, 2069 (1997),
hep-ph/9704301;\\
 Yu.A.Simonov and J.A.Tjon, Phys. Rev.  {\bf D
62}, 014501 (2000), ibid {\bf D62},  094511 (2000).

\bibitem{5}
 Yu.A.Simonov, QCD and Topics in Hadron Physics, Lectures at
the       XVII International School of Physics, Lisbon, 29
September -4
       October, hep-ph/9911237

\bibitem{6} V.I.Shevchenko and Yu.A.Simonov,
     Phys. Lett. {\bf B437}, 146 (1998) \\L.Lukaszuk, E.Leader and A.Johansen, Nucl. Phys. {\bf B562}, 291 (1999)\\
      original formulaton see in
     S.V.Ivanov, G.P.Korchemsky, Phys. Lett. {B154},  197 (1989)\\
     S.V.Ivanov, G.P.Korchemsky, A.V. Radyshkin, Sov J. Nucl. Phys.
      {\bf 44},  145 (1986).


\bibitem{7}
 V.I.Shevchenko and Yu.A.Simonov, Phys. Rev. Lett. {\bf 85}, 1811 (2000),
 hep-ph/0001299,
hep-ph/0104135


\bibitem{8} G.S.Bali, Phys. Rev. {\bf D 62}, 114503 (2000);\\
S.Deldar, Phys. Rev. {\bf D 62}, 034509 (2000).

   \bibitem{9}
L.Del Debbio, A. Di Giacomo and Yu.A.Simonov, Phys. Lett.
      {\bf B332}, 111 (1994);\\D.S.Kuzmenko and Yu.A.Simonov,
      Phys. Lett.  {\bf B494}, 81 (2000).

\bibitem{10}
M.Gell-Mann, R.L.Oakes and B.Renner, Phys. Rev. {\bf 175}, 2195
(1968);\\ S.L.Glashow and S.Weinberg, Phys. Rev.  Lett. {\bf 20},
224 (1968).


\bibitem{11}
     Yu.A.Simonov,
     Yad. Fiz.  {\bf 57}, 1491 (1994)

 \bibitem{12} A.M.Badalian and Yu.A.Simonov, Phys. Atom.
Nucl. {\bf 60}, 630 (1997).

  \bibitem{13}
A.M.Badalian and B.L.G.Bakker, Phys. Rev. {\bf D66} 034025 (2002)

\bibitem{14}

A.M.Badalian, B.L.G.Bakker and Yu.A.Simonov, Phys. Rev. {\bf D66}
034026 (2002) hep-ph/0204088


\end{thebibliography}
\end{document}